\documentclass{article}

\usepackage{arxiv}

\usepackage[utf8]{inputenc} % allow utf-8 input
\usepackage[T1]{fontenc}    % use 8-bit T1 fonts
\usepackage{hyperref}       % hyperlinks
\usepackage{url}            % simple URL typesetting
\usepackage{booktabs}       % professional-quality tables
\usepackage{amsfonts}       % blackboard math symbols
\usepackage{nicefrac}       % compact symbols for 1/2, etc.
\usepackage{microtype}      % microtypography
\usepackage{lipsum}		% Can be removed after putting your text content
\usepackage{graphicx}
\usepackage[numbers]{natbib}
\usepackage{doi}

\title{Word frequency and sentiment analysis of twitter messages during Coronavirus pandemic}

\author{Nikhil Kumar Rajput \\
  Department of Computer Science\\
  Ramanujan College (University of Delhi)\\
  India \\
  \texttt{n.rajput@ramanujan.du.ac.in}\\
	\And
	 Bhavya Ahuja Grover \\
  Department of Computer Science\\
  Ramanujan College (University of Delhi)\\
  India \\
  \texttt{b.ahuja@ramanujan.du.ac.in}\\
  \And
  Vipin Kumar Rathi \\
  Department of Computer Science\\
  Ramanujan College (University of Delhi)\\
  India \\
  \texttt{vipkrathi2013@gmail.com}\\
 \And
 Riya Bansal \\
 Department of Computer Science\\
  University of Delhi\\
  India \\
  \texttt{rbansal1@cs.du.ac.in}\\
 }
\renewcommand{\shorttitle}{}

\hypersetup{
pdftitle={A template for the arxiv style},
pdfsubject={q-bio.NC, q-bio.QM},
pdfauthor={David S.~Hippocampus, Elias D.~Striatum},
pdfkeywords={First keyword, Second keyword, More},
}

\begin{document}
\maketitle

\begin{abstract}
	The COVID-19 epidemic has had a great impact on social media conversation, especially on sites like Twitter, which has emerged as a hub for public reaction and information sharing. This paper deals by analyzing a vast dataset of Twitter messages related to this disease, starting from January 2020. Two approaches were used: a statistical analysis of word frequencies and a sentiment analysis to gauge user attitudes. Word frequencies are modeled using unigrams, bigrams, and trigrams, with power law distribution as the fitting model. The validity of the model is confirmed through metrics like Sum of Squared Errors (SSE), R-squared ($R^2$), and Root Mean Squared Error (RMSE). High $R^2$ and low SSE/RMSE values indicate a good fit for the model. Sentiment analysis is conducted to understand the general emotional tone of Twitter users messages. The results reveal that a majority of tweets exhibit neutral sentiment polarity, with only 2.57\% expressing negative polarity.
\end{abstract}

% keywords can be removed
\keywords{Word Frequency \and Power Law \and Coronavirus \and Social Media \and Twitter \and Sentiment Analysis}

\section{Introduction}
The year 2020 etched itself into history as a year when the world grappled with a formidable foe - the Coronavirus pandemic. This enormous disaster changed lives all throughout the world, cutting across national boundaries. Its tentacles reached far beyond the immediate realm of physical health, casting long shadows on economic stability, societal norms, and most profoundly, the human psyche.

The profound psychological impact of the pandemic is increasingly evident, prompting the exploration of innovative methods to capture and analyze human emotions. Platforms of social media, particularly Twitter and Facebook, offer unique windows into our collective psyche. Here, people share a kaleidoscope of emotions, weaving narratives of facts, fears, statistics, and dominant thoughts. This paper aims to shed light on this dynamic social media landscape, delving into the textual content that currently swirls around the pandemic through a statistical lens. We embark on a journey to unveil the intricate tapestry of human emotions woven during this unprecedented time, focusing specifically on the emotions/feelings expressed on the dynamic platform of Twitter. By diving into the vast ocean of tweets, we gain a unique perspective on the collective consciousness grappling with the pandemic's complex and far-reaching consequences.

To shed light on this terrain, two primary perspectives are utilized: word frequency analysis \cite{baayen2001word} and sentiment analysis \cite{taboada2016sentiment}. In quantitative linguistics, word frequency analysis is a well-known method for figuring out how frequently a word appears in a particular text or collection of texts. It assists in determining the most commonly used terms as well as their relative significance in that context. Using the power law \cite{clauset2009power}, a statistical tool proven to be effective in capturing such linguistic patterns, this study reveals patterns in the textual analysis by examining the probability distribution of word frequencies extracted from Twitter messages during the 2020 Coronavirus outbreak.

Sentiment analysis is performed in conjunction with this to determine the text's underlying emotional tone. This technique aids in deciphering attitudes towards specific subjects, providing a nuanced understanding of the emotional content within the discourse. As an intriguing and highly relevant field of research, sentiment analysis enriches our ability to infer and comprehend the emotional dimensions encapsulated in the textual expressions during the Coronavirus pandemic. This research seeks to provide important insights into the language nuances and collective attitudes that are present in the Twitter conversation during the global health crisis by merging different analytical methodologies.

There has been an increasing amount of interest in the use of social media data to analyze public opinions and conversations during the COVID-19 outbreak. Studies have indicated that Twitter data in particular can offer important insights into how the public is reacting to the pandemic. For instance, a study by Lwin et al. \cite{lwin2020global} demonstrated the analysis of global sentiments surrounding the COVID-19 pandemic on Twitter, highlighting the platform's potential for understanding public trends and attitudes. Additionally, another work \cite{pandey2022understanding, amores2023conversation} focused on the conversation around COVID-19 on Twitter during the first wave of the pandemic, emphasizing the application of sentiment analysis and topic modeling to analyze tweets published in English. People's opinions about COVID-19 were subjected to sentiment analysis by Kaur and Sharma \cite{kaur2020twitter}. They collected relevant tweets using the Twitter API, analyzed positive, negative, and neutral sentiments using techniques of machine learning, and pre-processed tweets using the NLTK package. The Textblob dataset served as the basis for the analysis, and the results were displayed using a variety of visualizations that highlighted neutral, positive, and negative opinions. Another study by Umer et al. \cite{umer2022etcnn} used ensemble model which is a combination of machine learning and deep learning models and uses the advantages of manually created features with automatic feature extraction. In this, TextBlob and VADER was used, unstructured data is collected, preprocessed, and analyzed prior to machine learning model training. In the same way, the effectiveness of the Word2Vec, TF, and TF-IDF features was examined. The outcomes showed that machine learning models work better using TextBlob and TF-IDF. 

A study by Sunitha et al. \cite{sunitha2022twitter} examined the real-time coronavirus-related tweets using a sentiment analysis approach. Originally, tweets from approximately 3100 European and Indian users were gathered between March 23, 2020, and November 1, 2021. For a deeper comprehension of the gathered data, pre-processing and exploratory research were then completed. Additionally, GloVe, pre-trained Word2Vec, Term Frequency-Inverse Document Frequency (TF-IDF), and quick text embeddings were used to achieve the feature extraction. The acquired feature vectors were then supplied to the ensemble classifiers, which included the GRU and the CapsNet neural network that categorized the user's emotions into four categories: fear, joy, sadness, and rage. The experimental results collected demonstrated that the suggested model was able to classify the sentiments of both Indian and European people with prediction accuracy of 97.28\% and 95.20\%, respectively. A study by Vijay et al. \cite{vijay2020sentiment} examined Indian tweets on COVID-19 (Nov 2019-May 2020) categorized as positive, negative, or neutral. Statewise, monthly, and overall datasets revealed initial negativity shifted to positive by April 2020, with focus on overcoming the virus. These studies underscore the significance of leveraging Twitter data and natural language processing techniques to gain a deeper understanding of public discourse and sentiments during the COVID-19 outbreak. 

The study is organized as follows: section \ref{word_frequency} includes word frequency analysis and the emergence of power law along with a brief summary of existing research in this area, section \ref{statistical_analysis} includes a statistical analysis of tweets, section \ref{sentiment_analysis} presents an overview of sentiment analysis and, section \ref{Methodology} includes an adopted methodology for performing sentiment analysis. The results of analyzing the sentiment of these Twitter messages are presented in section \ref{results} and the conclusion of paper is in section \ref{conclusion}.

\section{Word frequency analysis and power law} \label{word_frequency}

Word frequency analysis explores the numerical world of text to ascertain the frequency at which a given word occurs. It illuminates important themes and patterns in the examined data by identifying the most prevalent and recurrent phrases through counting occurrences and generating frequency distributions. Power law, a mathematical rule describing connections between changing quantities, finds its way into word frequency analysis too. It often governs the distribution of word frequencies, predicting that a few words will be used much more than most, creating a ``few common, many rare" pattern prevalent in natural language and text corpora. In the context of textual analysis, power law suggests that a few words (often referred to as ``stop words" like ``the", ``and", ``is") occur very frequently, while the majority of words occur rarely.

Many investigators have devised statistical and mathematical techniques to assess literary artifacts. A significant method is inferring the pattern of frequency distributions of every word in the content \cite{baayen2001word}. Zipf's law is mostly inherent in word frequency distributions \cite{zipf1935psycho,li1992random}. The law states that for a word vector $x$, the word frequency distribution $\nu$ changes as an inverse power of $x$. Other widely used distributions are Zipf-Mandelbrot \cite{mandelbrot1965information}, lognormal \cite{baayen1992statistical,carroll1968word}, and Gauss-Poisson \cite{baayen1992statistical}. Research similar to this has been conducted in several languages, including Hindi \cite{jayaram2008zipf}, Chinese \cite{shtrikman1994some}, Japanese \cite{miyazima2000power}, and many more \cite{baayen2001word}. Studies on single- and multi-word frequencies have been carried out extensively. One example is \cite{solso1979bigram}, which studied the frequencies and versatilities of bigrams and trigrams and reported 577 distinct bigrams and 6,140 different trigrams.

The power law distribution is one among the most well known. Because of its peculiar characteristics, this ``non-normal" distribution has drawn an extensive amount of interest in the academic world. The following is the mathematical representation of the rightly skewed distribution:
\begin{equation}\label{pl}
  f(x)=ax^b
\end{equation}
where a is constant and b is the scaling or exponential parameter.

Power law has been used in several research projects. The authors of \cite{srinivas2016link} highlight the existence of power law in social networks and make advantage of this characteristic to develop a degree threshold-based similarity metric that aids in link prediction. The authors assert that the power spectrum of the departure process resembles a power law comparable to that of observed traffic during the fragmentation of the data into Ethernet frames, in an attempt to describe the self-similar computer network traffic. They further claim that the input procedure had no bearing on the power law \cite{field2004measurement}. Power law modeling of internet topologies was demonstrated in \cite{faloutsos1999power}. The authors demonstrated that 5 out of 24 term frequency distributions and query frequency could be best fitted by a power law while looking into the existence of power laws in information retrieval data \cite{petersen2016power}. Power law is incredibly useful in many different fields. In this paper, we intend to use it to model the word frequencies of twitter messages posted during this time.

\section{Statistical analysis of tweets} \label{statistical_analysis}

This section contains the particulars of the study we conducted on the data we collected about tweets posted on Twitter from January 2020 to the present, which is the period of time after the global media began reporting on the coronavirus outbreak in China. The word frequency data collected from \cite{banda_juan_m_2020_3757272} corresponds to the tweets. According to the data source, there were about more than 4 million tweets every day starting on March 11, 2020, as awareness increased. Also, the data prominently captures the tweets in English, Spanish, and French languages. 

\begin{figure}
    \centering
    \includegraphics[width=0.85\linewidth]{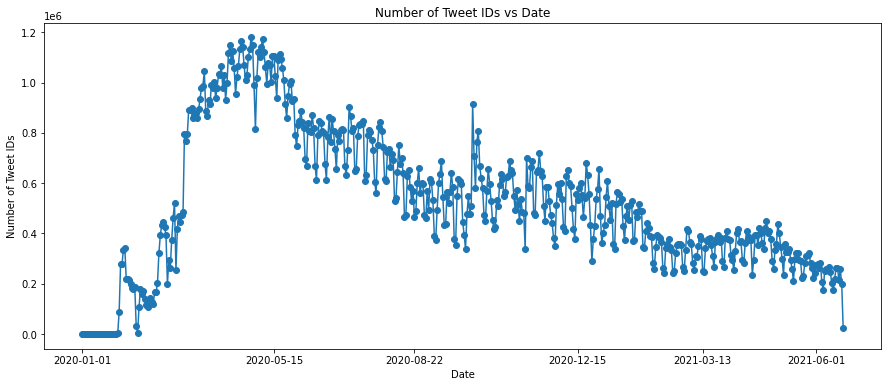}
    \caption{Evolution of number of Twitter ids involved in covid-19 posts}
    \label{fig:tweetid}
\end{figure}

A total of four analysis have been done to analyze the study. The first is the data on Twitter Id evolution reflecting on number of tweets.  It focuses on how many distinct Twitter IDs there were during a certain time frame that tweeted about the coronavirus. Figure \ref{fig:tweetid} illustrates the associated data and displays the trend of user activity over time. It demonstrates that the number of evolving Twitter IDs peaked in the months of March and April.

\begin{figure}
    \centering
    \includegraphics[width=0.5\linewidth]{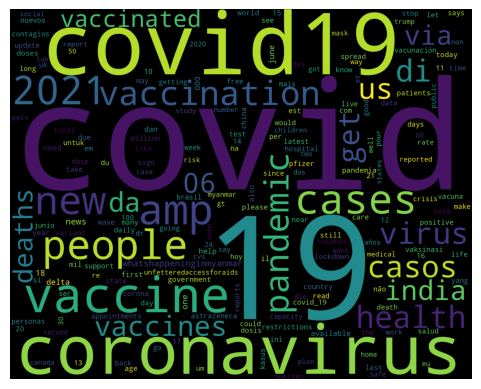}
    \caption{Unigram word cloud}
    \label{fig:wcloud}
\end{figure}

Fig. \ref{fig:wcloud} displays a word cloud visualization that highlights the most commonly appearing term, ``covid." It suggests that the public is highly engaged in the pandemic and the term is frequently used by Twitter users in their tweets. It can be inferred that people were tweeting about cases, news, or personal experiences. Also, the word ``covid" is a more direct and efficient way to say ``COVID-19". Its dominance indicates that consumers valued conciseness and speed above length while tweeting.

\begin{figure}
    \centering
    \includegraphics[width=0.6\linewidth]{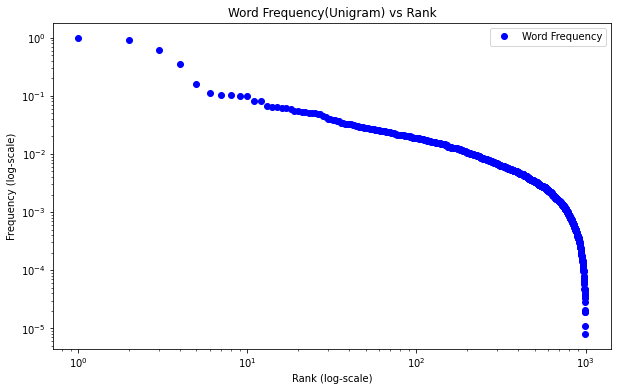}
    \caption{Plot for Unigram Frequencies vs Rank}
    \label{fig:unigram}
\end{figure}

\begin{figure}
    \centering
    \includegraphics[width=0.6\linewidth]{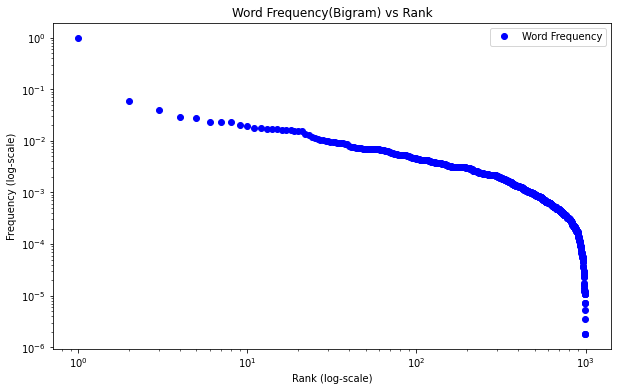}
    \caption{Plot for Bigram Frequencies vs Rank}
    \label{fig:bigram}
\end{figure}

\begin{figure}
    \centering
    \includegraphics[width=0.6\linewidth]{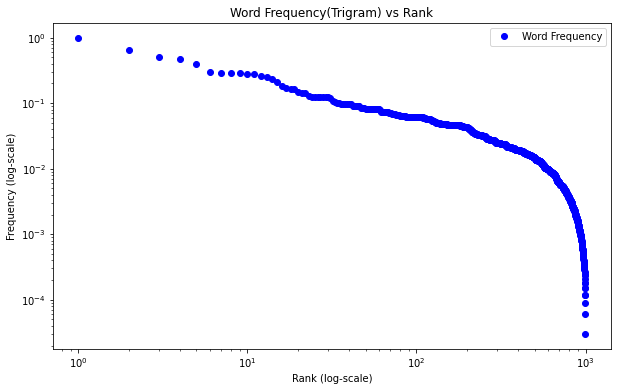}
    \caption{Plot for Trigram Frequencies vs Rank}
    \label{fig:trigram}
\end{figure}

The other three are unigram, bigram and trigram frequencies of words. The study examined the frequency of single-word combinations (unigrams), two-word phrases (bigrams), and three-word sequences (trigrams) in order to get insight into the world of Twitter. For this analysis, they concentrated on the top 1,000 examples of each kind. The relationship between rank (or index) and frequency for unigrams, bigrams, and trigrams is shown in the frequency distribution plots in Figs. \ref{fig:unigram}, \ref{fig:bigram}, and \ref{fig:trigram}, respectively. A power-law distribution is seen in these graphical representations, suggesting that the data and the power-law model fit the data quite well.

The calculated exponents for unigrams, bigrams and trigram are -0.9305, -3.7187 and -0.6514, respectively. The corresponding parameters are reported in Table \ref{table:tb1}. Notably, heavy tails are observed, particularly in the cases of unigrams and trigrams. The effective fit by the power-law distribution suggests a distinct behavior in tweet messages compared to literary documents like novels and poems, which adhere to Zipf's law with exponents close to 1.

% \subsection{Evaluating the Goodness of fit of the model}

Three goodness of fit metrics have been used to assess how well the power law distribution fits the data: SSE, $R^2$, and RMSE. The value obtained for the three datasets with the three forms of token of words has been shown in Table \ref{table:tb1}. We observe a notably elevated $R^2$ value across all three scenarios: unigram (0.9158), bigram (0.9863), and trigram (0.9764). Furthermore, the obtained values for SSE and RMSE are considerably minimal in each case. These results affirm the suitability of the power-law model for representing the frequency distribution of tweet message data.

\begin{table}[h]
\caption{Parameter values for Power Law distribution and goodness of fit}
\centering
\begin{tabular}{llll}
\hline
                & \textbf{Unigram}  & \textbf{Bigram}   & \textbf{Trigram}  \\ \hline 
\textbf{Parameters}      &          &          &          \\ 

% \hline
a               & 1.1450   & 0.9994  & 1.0566  \\ 
% \hline
b               & -0.9305   & -3.7187   & -0.6514  \\ 
\hline

\textbf{Goodness of fit} &          &          &          \\ 

% \hline
SSE             & 0.2063 & 0.0137 & 0.0750 \\ 
% \hline
$R^2$              & 0.9158   & 0.9863   & 0.9764   \\ 
% \hline
RMSE            & 0.0143  & 0.0037 & 0.0086 \\ \hline
\end{tabular}
\label{table:tb1}
\end{table}

\section{Sentiment Analysis of Twitter Messages} \label{sentiment_analysis}

Sentiment analysis is a fast growing field due to its capacity to interpret emotional quotient of a text.  A common definition of it is a computational investigation into people's attitudes, feelings, and views toward a topic \cite{medhat2014sentiment}. Sentiment analysis is mostly used to evaluate viewpoints, uncover latent emotions, and finally to classify their polarity into negative, positive or neutral. Some examples related to applications for sentiment analysis include customer reviews \cite{kang2012measuring}, news and blogs \cite{godbole2007large}, and the stock market \cite{nguyen2015topic}. A number of techniques, such as Artificial Neural Networks \cite{moraes2013document}, Naive Bayes \cite{rui2013whose} and, Support Vector Machines \cite{chen2011quality} have been used for sentiment analysis. A number of publications have also offered algorithms for sentiment analysis on tweets \cite{agarwal2011sentiment}, \cite{ghiassi2013twitter}, \cite{severyn2015twitter}, \cite{saif2012semantic}.

\section{Methodology} \label{Methodology}

\begin{figure}
    \centering
    \includegraphics[width=1\linewidth]{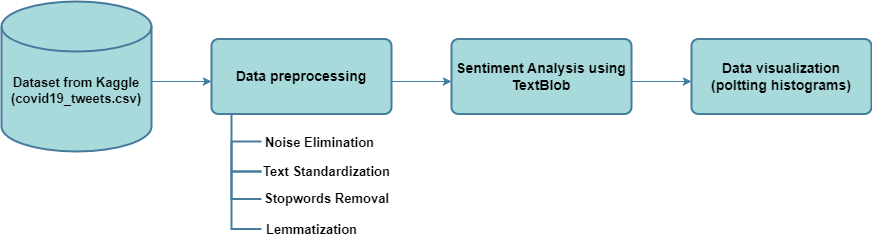}
    \caption{Workflow of sentiment analysis of COVID19 tweets}
    \label{fig:methodology}
\end{figure}

The methodology employed for sentiment analysis on a dataset of tweets about the COVID-19 epidemic is described in this section. Data collecting, data preparation, sentiment analysis, and data visualization are the four primary steps of the entire process. The process flow is illustrated in Fig. \ref{fig:methodology}.

\subsection{Data Collection}

Data has been collected from the COVID19 Tweets Kaggle dataset, which can be accessed at \url{https://www.kaggle.com/datasets/gpreda/covid19-tweets}. The dataset was gathered through the use of a Python script and the Twitter API, and it consists of tweets that come from various geographical places. To accumulate a substantial number of tweet samples, a daily query for the prominent hashtag \#covid19 has been executed for a specific duration, starting from July 25, 2020. The initial batch consisted of 17,000 tweets, and the collection process continued on a daily basis. The dataset is characterized by its focus on tweets related to the Covid-19 pandemic, particularly those featuring the \#covid19 hashtag.

\subsection{Data Preprocessing}

Since raw text data can contain insignificant information that can hinder accurate sentiment analysis, the collected tweets underwent a crucial preprocessing stage to refine them for further analysis. It is carried out with Python's NLTK (Natural Language Toolkit) module \cite{rajput2020natural} utilizing Natural Language Processing (NLP) approaches \cite{rajput2020natural}. This stage involved various techniques to clean and standardize the text data. The following used are noise elimination, text standardization, stop words removal and lemmatization.

\subsubsection{Noise Elimination}
The initial preprocessing step concentrated on eliminating irrelevant information that minimally impacts sentiment analysis. This ``noise removal" process targeted elements like URLs, mentions (@usernames), hashtags, special characters, and numbers. These elements were excluded because they don't significantly contribute to understanding the emotional tone of the tweets. To achieve this efficiently, Python's ``re" module was utilized. This module makes it possible to remove these unnecessary components from the gathered tweets by enabling the use of regular expressions, which are effective tools for pattern-based searching and removal. Further preprocessing procedures build upon this cleaning process, guaranteeing that the most relevant elements of the text input are kept in mind for precise sentiment analysis.

\subsubsection{Text standardization}
Following the initial noise removal, the text underwent a standardization stage to ensure consistency and eliminate potential sources of confusion for the sentiment analysis process. This standardization involves two key steps:
\begin{enumerate}
    \item \textbf{Punctuation Removal:} Punctuation marks, such as commas, periods, and exclamation points, were eliminated using Python's ``translate" function. This simplifies the text by removing elements that don't directly contribute to understanding the underlying sentiment. 
    \item \textbf{Whitespace Trimming:} Leading and trailing whitespaces, including extra spaces at the beginning or end of the tweet were removed using the ``strip" method. This ensures the text starts and ends cleanly, without unnecessary characters that might affect the model's interpretation. By eliminating these inconsistencies, the standardization process contributes to a more uniform dataset, allowing the sentiment analysis model to focus on the essential content of the tweets.
\end{enumerate}

\subsubsection{Removal of stop words}
After the text was standardized, the next step aimed to further refine the data by focusing on the words themselves. This involved stop word removal, a process that eliminates common words with minimal impact on sentiment analysis. Examples of stop words include ``the", ``a", ``is" and ``and." While these words serve grammatical functions and contribute to the overall meaning of a sentence in everyday language, they often hold little value in understanding the emotional tone of a tweet.

The NLTK library \cite{bird2009natural} in Python was utilized for this task. This library provides a readily available list of stop words in various languages, allowing for efficient removal of these common words from the collected tweets. By eliminating stop words, the focus shifts towards the words that carry more emotional weight, ultimately leading to a more accurate understanding of the sentiment expressed in the tweets.

\subsubsection{Lemmatization} 
Lemmatization is a technique in NLP used to reduce words to their base form, also known as the lemma \cite{khyani2021interpretation}. Unlike stemming, which focuses solely on the words stem, lemmatization takes context and grammatical role into account to ensure the resulting base form is a meaningful word.  For instance, ``running" becomes ``run", ``changing/changes/changed" becomes ``change". This ensures that different grammatical variations of the same word are treated consistently, allowing the sentiment analysis model to accurately capture the overall sentiment regardless of the specific word form used. NLTK provided the WordNetLemmatizer() function within its Python library and was employed to lemmatize the text data.

\subsection{Sentiment Analysis using TextBlob}

Following the data preprocessing stage, the sentiment analysis delves into uncovering the emotional undertones of the collected tweets. This step aims to quantify the overall sentiment, assigning a numerical value called polarity to each tweet. We utilize the TextBlob library \cite{loria2018textblob}\cite{Tutorial:-2024-03-02} in Python for this task. TextBlob offers a polarity score ranging from -1 (highly negative) to +1 (highly positive). Based on the calculated polarity scores, the tweets are categorized into specific sentiment classes:

\textbf{Positive:} Tweets with scores in the range (0, 1] are classified as positive. 

\textbf{Negative:} Tweets with scores in the range [-1, 0) are classified as negative.

\textbf{Neutral:} Tweets with a score of exactly 0 are classified as neutral.

\subsection{Data Visualization}
Histograms are a powerful tool for visualizing the distribution of data, making them well-suited for sentiment analysis of the collected COVID-19 tweets. Here we created a histogram with the x-axis representing the polarity scores (ranging from -1 to 1) and the y-axis representing the number of tweets/frequency within each score range. This visualization reveals the overall distribution of sentiment across the dataset, indicating whether the majority of tweets lean towards positive, negative, or neutral sentiment.

\section{Results of Sentiment Analysis of Twitter Messages} \label{results}

\begin{figure*}
    \centering
    \includegraphics[width=0.6\linewidth]{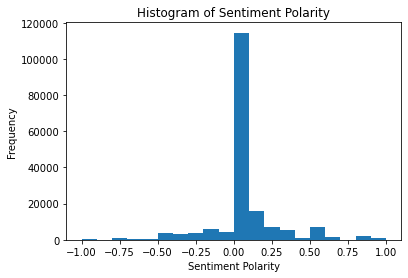}
    \caption{Histogram of sentiment polarities of tweets on Coronavirus}
    \label{fig:tot1}
\end{figure*}

\begin{figure*}
    \centering
    \includegraphics[width=0.6\linewidth]{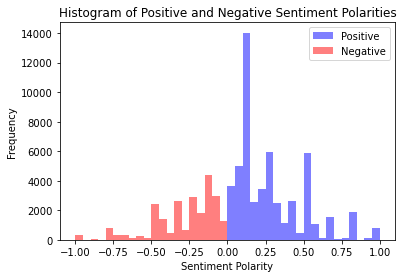}
    \caption{Histogram for positive and negative polarity of tweets}
    \label{fig:tot2}
\end{figure*}

To illustrate the general emotion, polarity values have been shown in a histogram (i.e., more positivity or negativity). Figs. \ref{fig:tot1} and \ref{fig:tot2} display the graphs. Based on the polarities in the dataset, Table \ref{table:tb2} shows the percentage of positive, negative, and neutral tweets. 

\begin{table}[h]
\caption{Polarities of sentiment analysis of tweets}
\centering
\begin{tabular}{llll}
\hline
\textbf{Sentiment Polarity}                & \textbf{Positive}  & \textbf{Neutral}   & \textbf{Negative}  \\ \hline

Tweets               & 6.45$\%$   & 90.97$\%$  & 2.57$\%$  \\ \hline
% Tweets by general public               & 29.33$\%$   & 54.92$\%$   & 15.75$\%$  \\ \hline

\end{tabular}
\label{table:tb2}
\end{table}

Fig. \ref{fig:tot1} corresponds to the histogram of sentiment polarities of tweets on COVID19 Tweets. It can be seen that majority of the tweets have a neutral sentiment followed by positive. The same can be inferred from Table \ref{table:tb2} that shows that around 90.97$\%$ tweets are neutral, 6.45$\%$ positive and a mere 2.57$\%$ is negative.

Fig. \ref{fig:tot2} represent the histograms produced by removing the neutral tweets. It readily reiterates that the positive emotions in the tweets are higher than negative ones. It shows that humans still post positive tweets and focuses on positivity rather than negativity.

\section{Conclusion} \label{conclusion}

This work explores the statistical analysis of tweets posted during the COVID-19 outbreak. As the pandemic swept across the globe, it left a trail of fear, uncertainty, and potential risks in its wake. This study seeks to analyze how these anxieties and the overall situation were reflected in the content of tweets during that time. To examine the COVID19 tweets, two methods have been employed: word frequency analysis and sentiment analysis.  According to the statistics, messages are more prevalent in the months of March and April 2020, which suggests how tweets have changed during that time. A word frequency analysis revealing the frequency of occurrences of each word in the tweets has been conducted. The top three terms with the highest frequency of occurrence in user-generated tweets are covid, 19, and covid19. For the top 1,000 instances, unigram, bigram and trigram frequencies were plotted and all of the plots fits the rightly skewed power law distribution. A heavy tail was observed in the plots of unigram and bigram. The exponential parameters obtained were -0.9305, -3.7187 and -0.6514 for unigram, bigram and trigram respectively. The model was validated through metrics like SSE, $R^2$ and RMSE. High value of $R^2$ and low value of SSE and RMSE were obtained which indicated a good fit for the model. The tweet dataset was also subjected to a sentiment analysis, and the related sentiment polarity (positive, neutral, or negative) and histograms were plotted. The majority of tweets (90.97\%) had neutral polarity, followed by positive polarity (6.45\%), according to the statistics. It demonstrates that people's feelings toward the COVID19 scenario were more neutral.

\bibliographystyle{unsrt}
\bibliography{reference}  %%% Uncomment this line and comment out the ``thebibliography'' section below to use the external .bib file (using bibtex) .

%%% Uncomment this section and comment out the \bibliography{references} line above to use inline references.
% \begin{thebibliography}{1}

% 	\bibitem{kour2014real}
% 	George Kour and Raid Saabne.
% 	\newblock Real-time segmentation of on-line handwritten arabic script.
% 	\newblock In {\em Frontiers in Handwriting Recognition (ICFHR), 2014 14th
% 			International Conference on}, pages 417--422. IEEE, 2014.

% 	\bibitem{kour2014fast}
% 	George Kour and Raid Saabne.
% 	\newblock Fast classification of handwritten on-line arabic characters.
% 	\newblock In {\em Soft Computing and Pattern Recognition (SoCPaR), 2014 6th
% 			International Conference of}, pages 312--318. IEEE, 2014.

% 	\bibitem{hadash2018estimate}
% 	Guy Hadash, Einat Kermany, Boaz Carmeli, Ofer Lavi, George Kour, and Alon
% 	Jacovi.
% 	\newblock Estimate and replace: A novel approach to integrating deep neural
% 	networks with existing applications.
% 	\newblock {\em arXiv preprint arXiv:1804.09028}, 2018.

% \end{thebibliography}

\end{document}